%Paper: astro-ph/9308037
%From: Yoshiaki Sofue <sofue@sof.mtk.ioa.s.u-tokyo.ac.jp>
%Date: Fri, 27 Aug 93 15:09:52 JST

%to appear in AJ
%\magnification=1200
%%%%%%%%%%%%%%%%
%% Definitions:
%%%%%%%%%%%%%%%%
\def\sect{\vskip 2mm \centerline}

\def\r{\hangindent=1pc  \noindent}
\def\ref{\hangindent=1pc  \noindent}
\def\cen{\centerline}

\def\v{\vskip 1mm}

\def\endpage{\vfil\break}

\def\noi{\noindent}
\def\kms{km s$^{-1}$}
\def\micron{$\mu$m}
\def\deg{$^\circ$}

\def\Halpha{H$\alpha$}

\def\Deg{^\circ}
\def\deg{$^\circ$}

\def\pa{PASJ}

\def\apj{ApJ}

\def\aj{AJ}
\def\aa{A\&A}

\def\aas{A\&AS}

\def\araa{ARAA}

\def\pasp{PASP}

\def\so{Sofue, Y.}
\def\ha{Handa, T.}
\def\na{Nakai, N.}
\def\fu{Fujimoto, M.}

\noi(to appear in  AJ.) % 1993.5.7

\v

\centerline{\bf A POLAR-NULCEUS DARK LANE IN THE BARRED SPIRAL M83:}
\centerline{\bf THREE-DIMENSIONAL ACCRETION IN THE NUCLEUS}
\v\v
\centerline{Yoshiaki SOFUE}
\centerline{\it Institute of Astronomy,  University of Tokyo, Mitaka, Tokyo
181}\v
\centerline{Ken-ichi WAKAMATSU\footnote*{A guest observer at Las
Canpanas Observatory, Carnegie Institution of Washington.}}
\centerline{\it Department of Physics, Gifu University, Gifu 505-11}
\v
\v\centerline{(Received: 1993 \hskip 30mm)}
\v\v

\centerline{\bf ABSTRACT}
\v
The central region of the barred spiral galaxy M83 reveals a  polar-nucleus
dust
lane, which extends from the NE molecular bar and crosses the central bulge.
Its SW counterpart is not visible, being hidden behind the bulge.
This asymmetry, in spite of the galaxy's face-on orientation and the symmetric
bar structure in the CO-line emission, indicates that the dark lane is an
off-plane structure.
Such a ``polar-nucleus'' structure can be formed by a non-coplanar,
three-dimensional accretion in a warped disk.

{\bf Subject Headings}: Accretion; three-dimensional -- Dust lane --
Galaxies; barred -- Galaxies; individual (M83) --  Galaxies; nuclei.

\v\v

\v\cen{\bf 1. INTRODUCTION} \v

M83 (NGC 5236) is a typical barred spiral galaxy of SBc type, rich in gas and
dust, with an almost face-on orientation.
Optical photographs show prominent dark lanes along the bar (Fig. 1), and,
since the bulge size is small, the shocked gaseous lanes can be traced even
near the central region.
For its typical characteristics as a barred spiral, various theoretical
simulations of bar-shock accretion have modeled this galaxy
(e.g., S$\phi$rensen et al. 1976; Huntley et al. 1978).

In this paper, we report on an evidence for an off-plane, dense dust lane
in the polar region of the nuclues, and discuss the implication of a
three-dimensional accretion of bar-shocked interstellar gas in the central
region.

\v\cen{\bf 2. POLAR-NUCLEUS DUST LANE}\v

A photographic plate in B-band was taken with the Cassegrain camera of the 2.5m
duPont telescope at Las Campanas Observatory.
It was exposed on 103a-O emulsion through a filter Wratten 2C for 30 minutes,
and is shown in Fig. 1.
The bar-shocked dark lanes are clearly seen along the leading edges of the bar,
which extends in the NE-SW direction.
The central bulge is seen as the bright, round component near the center,
and the dark lanes reach the bulge from NE and SW.

The central region of M83 is enlarged in Fig. 2a, and an unsharp-masked image
of the same field is shown in Fig. 2b.
The cross in Fig. 3a marks the position of the near-IR nucleus (Gallais et al.
1991).
The most prominent feature found in Fig. 2b is the dense dust lane across the
 central bulge, which  runs in the north-south direction.
This dark lane is a smooth and bent extension of the dark lane along the
leading
edge of the NE bar.
On the other hand, the dark lane running along the SW bar apparently stops near
the southern edge of the central bulge, and cannot be seen across the central
region.
Although the dark lanes in the outer-main bar are symmetrically developed,
their central parts show significant asymmetry.
Note that the galaxy is nearly face-on with the inclination angle
of  $i=24\Deg $ (Comte 1981).

\cen{-- Fig. 1, 2, 3 --}

We may consider the following two possibilities for this apparent asymmetry of
the dark lanes:

(a) The dark lanes are co-planar features, but the southern lane does not
extend
toward the nuclear region, and stops near the edge of the bulge, having a
physically
asymmetric structure.

(b) The dark lanes are symmetric with respect to the nucleus, but the dark
lane coming from the NE bar is on the near side to us with respect to the
nulceus,
while the one from the SW bar runs behind the bulge.
Hence, a polar, circum-nucleus band is composed of the dark lanes.
In fact Fig. 2b reminds us of the dark band around the elliptical galaxy NGC
5128,
which is suggested to be a circum galactic interstellar gas (e.g. Sandage
1961).

In ordr to clarify which interpretation is more reasonable, we have referred to
molecular line observations.
In Fig. 3a we superpose a CO-line intensity map obtained with the Nobeyama mm
Array with an angular resolution of $12'' \times 6'' $ (Handa et al. 1993;
Ishizuki 1993) on the unsharp-masked $B$-band image.
This CO map, as well as the one obtained with the 45-m telescope at an angular
resolution of 16$''$ (Handa et al. 1990), shows that the central molecular bar
is  symmetric with respect to the center of the galaxy.
This implies that an equal amount of molecular gas, and therefore dust, is
present
in the NE and SW regions near the center.
This fact immediately denies possibility (a), and hence, we may conclude that
possibility (b) is more reasonable.

\cen{-- Fig. 3a, b --}

We call hereafter this structure a ``polar-nucleus (polar-bulge) dust lane'',
and illustrate its possible orientation in Fig. 4 schematically.
If the lane consists of a circum-nuclear circle of radius  $\sim8''$ [140 pc
for a distance of 3.7 Mpc (de Vaucouleurs 1979)], about equal to
the radius of the optical bulge as indicated in Fig. 4 (see also Fig. 3a),
its plane must be tilted from the glactic plane by about 70\deg.

\cen{-- Fig. 4 --}

Fig. 3b shows a velocity field of CO emission which was produced using channel
maps obtained by Ishizuki (1993).
This map, as well as the one obtained with the 45-m telescope (Handa et al
1990),
shows a significant deviation from a normal circular rotation:
(If the rotation were circular, the iso-velocity lines should run
perpendicularly
to the node, opening toward NW and SE directions as they leave from the nodal
line.)
The northern part of the polar dark lane shows a positive excess, while the
southern
CO emission part shows negative excess which will be associated with the
hidden (invisible) dark lane behind the bulge.
This velocity anomaly can be understood, if the polar dark lane is accreting
toward the nucleus at an inward velocity of a few tens of \kms:
the northern dark lane, which is in the near side of the bulge, is moving away
from us with positive velocity in addition to the rotation, and the southern
lane (invisible behind the bulge) is approaching us with negative
velocity (plus rotation).
Namely, the circum nuclear band of dark lanes as illustrated in Fig. 4 is
contracting in addition to its rotation.

\v\cen{\bf 3. DISCUSSION}\v

\v\noi{\it 3.1 Relation to the nucleus}\v

Gallais et al. (1991) made imaging observations with sub-arcsecond
resolution at $J$, $H$, and $K$-band, and detected a bright compact
source which is surrounded by an arc-like structure in the southwest.
{}From brightness and position on the two color $(J-H)$-vs-$(H-K)$ diagram,
they  identified this as the nucleus of the galaxy.
We measured its position by referring to knot \#6 on their Fig. 1.
Thereby, we assumed that their knot \#6 is the IR counterpart to the optical
knot as marked B in Fig. A1, which is the brightest optical knot in the western
side of the polar dark lane.
Knot B (and therefore IR-knot \#6) is with the 6-cm radio continuum peak (Cowan
and Branch's 1985; Condon et al 1982),  which is, however, shifted by
approximately $2(\pm 1)''$ ($\sim35$ pc) east to the optical knot.
We also stress that these sources show an excellent correlation with the
10\micron\ IR emission (Telesco 1988),
which suggests that knot B is an active star-forming region containing hot
dust.

The optical image of the nuclear region is apparently divided into two bright
regions by the heavy dust lane (polar-nuclear lane), and comprises many optical
knots.
There is no prominent knot in the $B$ band that should be identified
as the nucleus of the galaxy, and this is the reason why M83 is classified
as an amorphous nucleus galaxy by Sersic and Pastoriza (1967).
However, by a detailed inspection of the $B$-band image, we could identify
an optical knot as marked K in Fig. A1 with the position of the IR
nucleus (Appendix and Table A1) .
Radio continuum maps at 6 cm, as superposed on Fig. 3a,  also indicates
no particular source at the near-IR nucleus (Cowan and  Branch 1985; Condon et
al 1982).

\cen{-- Table 1--}

The obtained positions of the nucleus are listed in Table 1, and the IR
nucleus is marked in Fig. 3a by the cross.
The error in the here-determined position of the IR nucleus (IR knot \#1 of
Gallais et al 1991) is  $\pm 1.''5$.
Table 1 also lists peak positions of the CO line emission (Ishizuki 1993)
and radio continuum emission at 6 cm (Condon et al 1982).
Since CO emission is extended, its peak position is estimated to have an
error of $\pm 3''$, although the absolute positioning in the CO-line map (Fig.
3a)
from the Nobeyama mm-wave Array is better than $\pm 1''$,
As is readily seen in Fig. 3a, the near-IR nucleus is apparently deviated from
the
approximate center of the distributions of radio-continuum, molecular-gas, as
well as from the dynamical center (Ishizuki 1993), toward the NE by
about 3$''$ (50 pc).

\v\noi{\it 3.2. Circum-nulcear ring of star-forming regions}\v

Gallais et al. (1991) suggested that a ``near-IR arc'' (their Fig.  3) is
composed of a circumnuclear ring (or an ellipse) of active star forming
regions.
The ring can be identified with optical knots aligned along the SW edge of the
bright region in Fig. 2b, part of which is obscured by the dark lane.
This star-forming ring is asymmetric with respect to the nucleus, and its NE
counterpart is not visible even in the near-IR.
Such an apparent asymmetry could be explained if the ring is tilted with
respect to its node running at around PA$\simeq 130\Deg$ (approximately
perpendicular to the bar).
Such a tilted ring (ellipse) could be produced, if star formation took place in
a
dust lane (molecular gas band), which had accreted in the past in a similar
manner  to the present polar nuclear dust lane.

\v\noi{\it 3.3. Three-dimensional accretion in the nuclei}\v

Various theoretical studies have been made of accretion processes of
interstellar gas in a barred spiral galaxy (S$\phi$rensen et al. 1976;
Huntley et al. 1978; Fukunaga and Tosa 1991).
Along this guide line, efforts have been devoted to explain the feeding
mechanism of
AGN through bar-shoked accretion onto the nucleus (e.g. Noguchi 1988).
In these models, however, simulations have been made in a two-dimensional
scheme
under a ristricted condition that the interstellar gas is present in a flat
galactic plane.
Namely, if the gas is accreted toward the center, its flow inevitably
encounters the central region, but no way is allowed for the gas to escape
from the disk plane.
Such a ristricted condition might have resulted in an apparently efficient,
and therefore, artificial accretion to the nucleus.

If we take into account three-dimensional structures near the nucleus, the
efficiency of accretion to the nucleus might be reduced because of the increase
in the degree of freedom  in the $z$ direction.
In actual galaxies, galactic disks are known to be hardly co-planar,
but are more or less warped and corrugated.
This is becasue  the majority of galaxies are not isolated, but are
interacting.
Warping and corrugation occurs  easily by gravitational disturbances by
nearby galaxies as well as by interaction with the intergalactic gas.
In fact, observations have shown that the nodal line of the HI velocity field
of
M83 is significantly deviated (rotated) from a co-planar disk and showed that
the
HI disk is largely warped (Rogstad et al. 1974).
The velocity field of the inner disk of M83 shows also deviation from a
co-planar circular rotation (Fig. 3b; Handa et al. 1990; Ishizuki 1993),
and a part of such deviation could be due to a warp of the inner disk, although
separation from non-circular motion such as the contraction must not be easy.

We may conjecture the following for the formation of a polar-nucleus structure:
We assume that the galactic disk is not a flat plane, but is warped and
corrugated in the $z$-direction (perpendicular to the disk), which is really
observed in the outer HI disk (Rogstad et al 1974).
In the outer-main bar, the interstellar gas suffers from a shock compression at
leading edges of the bar due to interaction with the oval potential.
Hence, the azimuthal orbital angular momentum of the gas is lost, and the
gas is accreted toward the center.
However, the angular momentum due to the motion perpendicular to the plane
remains invariant during the accretion, since no shock occurs in the
$z$-direction.
Finally this angular momentum becomes dominant near the nucleus,
and produces a polar-nuclear ring of accreting gas.

We stress that such three-dimensional accretion is more realistic, and that,
if we take into account this, theoretical results so far obtained by
two-dimensional simulations would be significantly changed in the sense
that efficiency of accretion to the nucleus is reduced.

\sect{\bf APPENDIX}\v

In Fig. A1 we identified optical ($B$-band) knots in the central bulge of M83,
and indicated them by A to O.
We measured  their positions by referring to stars 1 to 5 in Fig. A1,
whose positions were measured by referring to SAO stars around
M83 (stars 6 to 10 in Fig. 1).
Coordinates of the knots and stars are given in Table A1.

\cen{-- Fig. A1, Table A1 --}

\v\v\cen{\bf References} \v

\r Comte, G. 1981, \aas, 44, 441.

\r Condon, J. J., Condon, M. A., Gisler, G., and Puschell, J. J. 1982, \apj,
252, 102.

\r{Cowan, J.J., and Branch, D. 1985, \apj, {\bf 293}, 400.}

\r de Vaucouleurs, G. 1979, \aj, 84, 1270.

\r{Fukunaga, M. and Tosa, M. 1991, \pa, {\bf 43}, 469. }

\r{Gallais, P., Rouan, D., Lacombe, F., Tiphene, D., and Vauglin, I. 1991,
\aa, {\bf 243}, 309}

\r \ha, Ishizuki, S., Kawabe, R. 1993, in {\it Astronomy with Millimeter and
Submillimeter Wave Interferometry, IAU Coll. No. 140}, ed. M. Ishiguro, in
press.

\r \ha, \na, \so,  M.Hayashi, and \fu 1990, \pa, {\bf 42}, 1.

\r{Huntley, J. M., Sanders, R. H., and Roberts, W. W.,  1978, \apj, {\bf 221},
521. }

\r Ishizuki, S. 1993, Ph.D. Thesis, University of Tokyo.

\r{Noguchi, M. 1988, \aa, {\bf 203}, 259. }

\r Rogstad, D. H., Lockhart, I. A., and Wright, M. C. H. 1974, \apj, 193, 309.

\r{Rumstay, K.S., and Kaufman, M. 1983, \apj, {\bf 274}, 611.}

\r Sandage, A. R.  1961, The Huble Atlas of Galaxies, Carnegie
Institution, Washington, p. 50

\r{S$\phi$rensen, S. -A., Matsuda, T., and Fujimoto, M. 1976,
 {\it Astrophys. Sp. Sci. }, {\bf 43}, 491. }

\r {Sersic, J.L., and Pastoriza, M.G. 1967, \pasp, {\bf79}, 152.}

\r Telesco, C. M. 1988, \araa, 26, 343.

\vskip 50mm

\settabs 7 \columns
\noindent{Table 1.  Positions of the nucleus at various wavelengths}
\v

\hrule \vskip 1mm \hrule
\v
\+  Object & R.A. & & Dec. (1950)  & Notes  \cr
\v

\hrule
\v

\+  IR nucleus   & 13h 34m 11.61($\pm0.11$)s & & $-29\Deg 36' 40(\pm1.5)''$ &&
Gallais et al. (1991); \cr

\+&& & & & =identified with optical knot K \cr

\+ \Halpha\ peak  & 13h 34m 11.55($\pm0.04$)s & & $-29\Deg 36'  42.2(\pm0.4)''$
&&
Rumstay and Kaufman (1983) \cr

\+  Peak of CO  & 13h 34m 11.4($\pm 0.2$)s  && $-29\Deg 36' 37(\pm 3)''$ &&
Ishizuki (1993); extended \cr

\+ & &&&& ($\sim$ kinematical center) \cr

\+  Radio peak & 13h 34m  11.10($\pm 0.01$)s &  & $-29\Deg 36'  35.2(\pm
0.3)''$ & &
Condon et al. (1982); 6 cm \cr

\+ &&&&& (may be not the nucleus) \cr

\v \hrule \v

\endpage

\settabs 9 \columns
\noindent{Table A1.  Positions of bright knots in the nuclear region of
M83 and of reference stars.$^\dagger$ }
\v

\hrule \vskip 1mm \hrule
\v
\+  Object & ~~~R.A. & & ~~~~Dec.&&  & Notes  \cr
\v

\hrule
\v

\+  Knot A  & ~~~13h 34m& 10.55($\pm0.11$)s   &
 ~~~~$-29\Deg 36'$& $38.4(\pm 1.5)''$ & \cr

\+  B  & &10.88s      & & 35.1 & knot \#6 in
Gallais et al (1991) \cr

\+   C  & &11.01s   &   & 49.2 & \cr

\+   D  & & 11.12s      &&41.9 &  knot \#5 \cr

\+  E  & & 11.24s      &&45.2 &  knot \#4?\cr

\+  F  & & 11.27s      &&38.7 & \cr
\+  G  && 11.29s      && 48.1 & \cr

\+  H  && 11.37s      && 43.6 &  knot \#4? \cr

\+  I  &&11.50s      && 40.6 &  knot \#7 \cr

\+  J  &&11.60s      && 46.0 &  knot \#3) \cr

\+  K  & & 11.67s      && 38.7 &  knot \#1; the nucleus \cr

\+  L  & & 11.68s      & &44.9 & \cr

\+  M  & & 11.68s      && 41.1 & \cr

\+  N  &&11.74s   &   & 46.5 &  knot \#2 \cr

\+  O  & & 11.89s  &    & 42.1 & \cr

\v

\+ Star 1  & ~~~13h 34m & 09.75($\pm0.04$)s      &
 ~~~~$-29\Deg 36'$& $09.9 (\pm 0.6)''$ & \cr

\+  2  && 12.62s      && 56.9 & \cr

\+  3  & &13.97s      && 23.8 & \cr

\+  4  & & 14.13s      && 59.3 & \cr

\+  5  & & 14.83s      && 59.2 & \cr

\v

\+  6 && 46.53s      && 00.5 & \cr

\+ 7 & & 58.78s      && 36.1 & \cr

\+ 8 & & 10.33s      && 26.3 & \cr

\+  9 & & 27.94s      & & 31.8 & \cr

\+  10 && 28.52s      && 50.7 & \cr

\v \hrule
\v
\noi $\dagger$Error in  positions are $1''.5$ for knots, and $0''.6$ for
stars in both coordinates. Knots A to O and stars 1 to 6 are indicated in
Fig. A1, and stars 6 to 10 in Fig. 1.

\vskip 20mm

\sect{Figure Captions}
\v\v\r Fig. 1:  A $B$-band image of M83. The line at the bottom-right corner
indicates $1'$.
Positions of stars 6 to 10 are listed in Table A1.

\v\r Fig. 2: (a) Left panel: The central region of M83 in $B$ band. The bar
indicates 30$''$.
(b) Right panel: The same field as (a), but the image has been
contrast-enhanced
by unsharp masking technique.
Note thea straight dark lane which crosses the bulge from the north to south.
The galaxy is nearly face on ($i=24\Deg$).

\v\r Fig. 3:
 (a) Superposition of contour maps of the CO-line intensity (Ishizuki et al.
1993)
and 6-cm radio continuum (Cowan and Branch 1985) on the same photograph as Fig.
2b.
The cross indicates the position of the IR nucleus (Gallais et al. 1991).

(b) CO-line velocity field for the same area as in (a), as derived from channel
maps presented by Ishizuki (1993). Contour numbers are in \kms.
The straight line shows a nodal line of the overall optical disk.

\v\r Fig. 4: Schematic illustration of the polar-nuclear (polar-bulge) dust
lanes (thick lines) and the circum-nuclear ring of active star
formation (small circles).

\v\r Fig. A1: Optical knots in $B$ band image. Their positions are given in
Table A1 together with those for reference stars 1 to 5.

\bye